\documentstyle[aps]{revtex} 
 
\begin{document} 
 
\title{The Measurement Problem and the Reduction Postulate  
of Quantum 
Mechanics}

\author{Rodolfo Gambini \\ 
Instituto de F\'{\i}sica, Facultad de Ciencias, \\ 
Tristan Narvaja 1674, 11200 Montevideo, Uruguay}

\maketitle 
 
\begin{abstract} 
It is well known, that the causal Schr\"odinger evolution of a quantum
state is not compatible with the reduction postulate, even when
decoherence is taken into account. The violation of the causal
evolution, introduced by the standard reduction postulate distinguishes
certain  systems (as the measurement devices),
whose states are very close to statistical mixtures
(as the ones resulting from the process of decoherence). In these
systems, this violation has not any observable effect. In arbitrary
quantum systems, the transition from the initial density matrix to a
diagonal matrix predicted by the standard reduction postulate, would
lead  to a complete breakdown of the Schr\"odinger evolution, and
consequently would destroy all the predictive power of quantum
mechanics. What we show here, is that there is a modified version of the
postulate that allows to treat all the quantum mechanical systems on
equal footing. The standard reduction postulate may be considered as a
good approximation, useful for practical purposes, of this modified
version which is consistent with the Schr\"odinger evolution and via
decoherence with the experimental results. Within this approach, the
physical role played by the reduction postulate is as a tool for the
computation of relative probabilities and consequently for the
determination of the probabilities of consistent histories.

\end{abstract} 
 
\vskip 3cm

It is well known that when two quantum systems temporarily  
interact, the 
composite system is generally left, after the interaction,  
in an 
"entangled state".That means that the state vector is not  
just a product 
of the states vectors corresponding to each system. 
This raises difficulties if one of the systems is  
macroscopic, more 
specifically if it is a measuring apparatus. After the  
measurement a 
linear superposition of macroscopically different states of  
the 
measuring apparatus is obtained. 
 
It is well known that Bohr's solution to this problem is  
based on the 
following assumption: a macroscopic object is not subject  
to the  rules 
of quantum mechanics but only to classical physics. Even  
though this 
rule removes the difficulty, it raises \cite{O} a host of  
further 
questions: When is an object big enough to be considered as  
classical 
rather than quantum? What laws of physics describe the  
transition 
between the quantum world and the classical one? Is it  
logically 
consistent to impose classical physics when quantum  
mechanics find a 
difficulty? 
 
If one wants to view the quantum theory as the general  
dynamical theory 
of all the physical systems, and not just microsystems in  
the 
laboratory, difficulties seem to be even harder. One needs  
to explain 
why the dynamics expressed by the Schr\"odinger equation is  
suddenly 
rejected in some very special interactions called  
measurements. This is 
even more surprising if one takes into account that the  
same 
Schr\"odinger equation is applied to a host of macroscopic  
objects as 
long as they are not involved in measurement processes. 
 One could assume that reduction is associated with the
production of 
events, and therefore the breaking of the causal evolution 
is nothing 
but the inclusion of the information supplied by the event. 
But as it is 
well known and we shall discuss in what follows, the 
the standard reduction postulate implies that the breaking of the
causal evolution occurs {\em even when the information 
resulting from the
measurement is not taken into account.}

Decoherence \cite{HL}, \cite{Zu}, \cite{JZe} allows to  
explain 
the evolution of the density operator of a 
macroscopic object into an {\em almost} diagonal matrix,  
when only 
collective observables are taken into account. It can take  
place, for 
instance, in a measuring apparatus and destroy any {\em  
practical} 
possibility of observing a quantum superposition 
of distinct macroscopic events. However if quantum  
mechanics is exactly 
correct and the Schr\"odinger equation rigorously describes  
the dynamical 
evolution, the phase coherence is never completely lost.  
This is the 
core of the measurement problem that is not completely  
solved by 
decoherence: how to combine the observation of an event 
by an observer $O$ with the fact that the phase is never  
lost and 
therefore the information about the other branches is still  
present, even 
if not for $O$. 
 
If one assumes  
the 
universal evolution according to the Schr\"odinger  
equation, then, the 
reduction of the density operator from an almost diagonal  
matrix to a 
rigorously diagonal one, as it is required by the reduction  
postulate, 
cannot occur\cite{D'E}. But it is not so easy to get rid of  
the reduction 
postulate. In fact, if the reduction does not happen, then  
the events 
observed by $O$ cannot belong to exactly consistent  
histories but only 
satisfy the consistency requirement in some approximation.  
But if that 
is the case, what is the counterpart of the production of  
events in the 
formalism?  What is even more important, if reduction does  
not occurs 
when an event is produced, it is not even possible to  
assign 
probabilities to  alternative histories. On the other
hand, if one assumes that reduction takes place, even if it 
implies the breaking of the causal evolution, one needs to  
assume that it only occurs in quantum systems (as measurement 
devices) whose states  are very close to statistical mixtures 
(as the ones  resulting from a process of decoherence). In fact, the
occurrence of reduction in arbitrary quantum systems would  
imply the complete breakdown of the Schr\"odinger
evolution and would destroy all the predictive power of
quantum mechanics.

Here, I will show that there is a very natural modification  
of the 
reduction postulate consistent with the rest of the quantum  
formalism 
and with the experimental results that allows to solve the  
measurement 
problem. In other words, the new reduction postulate is  
always 
consistent with the Schr\"odinger evolution.  Within this
new approach events can occur in any quantum mechanical  
system. If they are observed then the reduced state will  
allow to describe the future probabilities. If they are  
microscopic quantum events that are not observed, the  
reduction of the quantum states due to the production of   
events does not preclude the use of the Schr\"odinger
equation to describe the evolution of the system.

Let us first recall the standard form of the reduction  
postulate. We 
consider a system $S$ associated with the Hilbert space  
${\cal H}$ 
whose state is given by the density operator $\rho$, which  
is a 
self adjoint positive definite operator with unit trace.  
Let $P_1$ be 
the projector to a certain subset ${\cal E}_1$ of the  
spectrum of an 
observable operator $A$. Then, according to standard  
quantum mechanics, 
the probability of obtaining one of the eigenvalues of $A$  
belonging to 
the subset ${\cal E}_1$  is given by 
 
\begin{equation} 
p_1= Tr[\rho P_1] 
\end{equation} 
 
and according to the reduction postulate, the corresponding  
density 
operator after the measurement is given by 
 
\begin{equation} 
\rho_1= {{P_1 \rho P_1}\over {Tr[\rho P_1]}} \label{usual} 
\end{equation} 
 
Thus if one wants to describe the state of the system after  
the 
measurement has taken place, without including the  
information about the 
particular outcome of the measurement, it will be given by  
the density 
operator 
 
\begin{equation} 
\rho_F=\sum_n { p_n \rho_n} = \sum_n {P_n \rho P_n}  
\label{final} 
\end{equation} 
 
If, for instance the system under consideration is the  
Zurek's1982 
bit-by-bit model of a spin $1/2$ \cite{Zu} particle coupled  
with a  
measuring apparatus, the density operator after the  
decoherence process is 
described by 
 
\begin{equation} 
\rho = 
\left(\begin{array}{cc}|a|^2&z(t)\\z^*(t)&|b|^2\end{array}\right).
\label{ro} 
\end{equation} 
 
where z(t) is a very small quantity. If the reduction of  
the state takes 
place,  according to Eq \ref{final}, it will take the form 
 
\begin{equation} 
\rho_F =  
\left(\begin{array}{cc}|a|^2&0\\0&|b|^2\end{array}\right). 
\end{equation} 
 
Even though the change from $\rho$ to $\rho_F$ would be  
"fantastically" 
difficult to detect, this change is a sign of the violation  
of the 
causal evolution described by the Schr\"odinger equation. A  
fundamental 
physical law cannot be accepted with the argument that it  
is very 
difficult to prove that is false. Furthermore this 
violation will be 
small, if events can only occur after a decoherence 
process. That is in 
a very particular subset of quantum systems whose states 
are very close 
to a statistical mixture. 
 
        We shall now explore the possibility of introducing 
modifications of the  reduction postulate rigorously  
consistent with the 
Schr\"odinger evolution. According to the new postulate the  
evolution of a 
state will be always described by the Schr\"odinger  
equation, but when an 
event is observed the information about the density matrix  
will be 
actualized. We will therefore be interested in first place  
in the systems 
where events are observed. These systems are the  
macroscopic measurement 
apparatus coupled with a quantum system. Their states will  
be described by 
a partial trace after decoherence has taken place. The  
density matrix, 
in the "pointer basis" associated with the macroscopic  
states, presents 
after decoherence  the well known dumping of the  
interference terms 
 
\begin{equation} 
|\rho_{ij}| / \rho_{ii} << 1 \;\; \hbox{and} \;\;  
|\rho_{ij}| / \rho_{jj} << 1 
\end{equation} 
 
A one parameter family of possible reduced states  
consistent with 
Schr\"odinger can be obtained as follows. If the  
measurement of the 
observable $A$ leads to a value lying in the set ${\cal  
E}_1$, the 
density operator after the measurement may be given by the  
self-adjoint 
unit trace matrix 
 
\begin{equation} 
\rho_1={1 \over {Tr[\rho P_1]}} 
\{ \textstyle{1/2}[\rho P_1+ P_1 \rho]-\lambda[\sum_{j \ne  1}
{ |\rho_{1j}|]P_1}+ \lambda\sum_{k \ne 1}{|\rho_{1k}|P_k} \}
\end{equation} 
 
where $\lambda$ is an arbitrary real constant.  For each  
value of 
$\lambda$ one has an alternative for the reduction  
postulate. 
If $\lambda$ is grater or equal to $1$, the density matrix  
is positive 
definite When the outcome of the measurement is unknown,  
the state of 
the system after the measurement has taken place is given  
by 
 
\begin{equation} 
\rho_F=\sum_n {p_n \rho_n} = \textstyle{1/2}\sum_n(\rho  
P_n+ P_n 
\rho)=\rho 
\end{equation} 
 
and therefore the reduction process does not imply any  
abrupt change in 
the density operator. In other words the reduction of the  
state and the 
Scroedinger evolution are now totally compatible. Notice  
that the form 
of the state after the measurement differs from the usual  
expression 
(\ref{usual}) by interference terms and therefore due to  
decoherence the 
modification of the reduction postulate is undetectable in  
the 
measurement process. 
 
In order to single out an alternative, the arbitrary  
constant 
$\lambda$ should be fixed by physical considerations. To do  
that we 
notice that the 
reduced state depends explicitly on the pointer basis and  
therefore is 
not well suited to describe generic events, it only allows  
to describe 
events taking place in a measuring apparatus. This is not
acceptable,
a reduction postulate should allow to assign probabilities 
to any kind 
of event as the standard postulate does. 
 
If one wants to include microscopic quantum events in the  
description 
and assign probabilities to histories, we would not dispose  
of a 
preferred pointer basis and therefore the only possible  
definition of the 
reduced state is the one obtained for $\lambda=0$. For this  
value of 
$\lambda$ the reduced density operator is not always  
positive definite, 
but, as we shall prove, it allows to assign 
probabilities to consistent families of histories and  
practice 
retrodiction. 
 
Reduction is the counterpart in the formalism of the  
production of 
events. In our approach, reduction can occur at any instant  
of time 
provided a "choice" or "branching" among different  
possibilities occur. 
That is, if $P$ is the projector  corresponding to the  
event, provided 
\begin{equation} 
0< p= Tr[\rho P] \label{pos}< 1 
\end{equation} 
If this condition is fulfilled, then the property  
associated with $P$ 
can be actualized and, in that case, reduction will take  
place. The density  
operator will take the form 
 
\begin{equation} 
\rho_P={1 \over {Tr[\rho P]}} 
\textstyle{1/2}[\rho P+ P \rho]= {1 \over {Tr[\rho P]}} 
\textstyle{1/2}[ P, \rho]_+ 
\end{equation} 
 
It is now possible to assign probabilities to histories. 
A family of histories will be consistent if for each set of  
projectors, 
of the observables under consideration, on disjoint  
domains, whose union 
at a given time, covers the whole spectrum \cite{Gr,Om,GH}, 
and for each union of sets, one can assign a probability  
$p$ such that 
 
\begin{equation} 
0 \le  p=Tr[P_n(t_n)[P_{n-1}(t_{n- 
1}),.....[P_1(t_1),\rho]_+...]_+] \le 1 
\end{equation} 
 
satisfying the positivity condition. 
Notice that consistent histories are automatically  
additive. If 
$P_i={P^1}_i+{P^2}_i$ then  $p=p^1+p^2$ , due to the  
linearity 
of the reduction postulate in the projectors. 
Contrary to what happens with the standard postulate, the  
main requirement 
is positivity while additivity is automatically satisfied 
 
Due to decoherence, the consistency condition for  
macroscopic 
events would be practically always fulfilled.   However,  
for some 
observables "fantastically" difficult to observe, events  
would not 
occur. For instance let us suppose that  one starts from the state given in
Eq[\ref{ro}] 
and the outcome of the first measurement is $s_z=+1/2$, and let us
allow the system to evolve in time in such a way that after a time t,
the spin is approximately aligned with the x direction.
Then near this direction will be a window of angle of the order of the
dumped interference terms (that is, of order $2^{-N}$, where $N$ is the number
of degrees 
of freedom of the environment)
such that a second measurement
on that direction will not lead to the production of events.
Experiences leading to such situations seem to be not feasible. They
would not only require a fine tuning in the rotation angle of order
$2^{-N}$ but also a polarizer able to distinguish among angles of this
order.
 
The density operator resulting from a macroscopic  
measurement will 
have some negative eigenvalues with absolute values of 
the order of the interference terms and therefore it could  
be called 
quasi-positive definite. Due to the unitarity of the  
Schr\"odinger 
evolution it is immediate to check that the quasi- 
positivity of the 
density operator resulting from a macroscopic measurement  
is conserved 
by the evolution. 
 
To conclude, we have proposed a modification of the  
reduction postulate. 
Due to decoherence, this postulate practically coincides  
with the usual 
one, leading after a measurement process to a state which  
differs from 
the standard projected state by terms proportional to the  
interference 
components  in the decohered matrix, and insures an  
evolution rigorously 
consistent with the Schr\"odinger equation. It allows to  
assign 
probabilities to histories and consequently to speak of  
quantum events. 
The reduction of the state in a macroscopic system  
corresponds to an 
actualization of the information available for the  
observers of the 
event \cite{Ro} while the evolution of the Universe is  
always 
described by the Schr\"odinger equation. 
The new postulate involves the use of non positive definite  
density 
matrices as intermediate steps in the calculation of  
probabilities for 
consistent histories. As we have stressed these matrices  
are never 
obtained as the outcome of a macroscopic measurement. In  
this approach 
the physical role played by the reduction postulate is as a  
tool for the 
computation of relative probabilities and consequently for  
the 
determination of the probabilities of consistent histories.
{} From a philosophical  point of view  one can
adopt two 
alternative realistic interpretations: 1) The state of the system
changes each time an event is produced. However a particular
observer 
that does not know what events have been produced is 
allowed to use the state obtained by evolving the initial 
state with the 
Schr\"odinger equation. 2) If one adopts the Many Worlds interpretation,
one could consider that the state of the Universe never collapse, and
the reduced states are useful tools to compute conditional probabilities
within each brunch of the Universe. Notice that, if one uses the usual
reduction postulate to compute conditional probabilities the predictions
are not, strictly speaking, compatible with the causal evolution of the
state of the Universe.

\end{document}